\documentclass[preprint,nofootinbib,aps,superscriptaddress,showpacs]{revtex4}
\usepackage{color,graphicx,shortvrb,epsfig}

\usepackage{graphicx}
\usepackage{dcolumn}
\usepackage{bm}

\usepackage{colordvi}
\usepackage{psfig,epsfig,amssymb,amsmath,latexsym}

\newcommand{\no}{\nonumber}
\def\lsim{\mathrel{\rlap{\lower4pt\hbox{\hskip1pt$\sim$}}
    \raise1pt\hbox{$<$}}}         
\def\gsim{\mathrel{\rlap{\lower4pt\hbox{\hskip1pt$\sim$}}
    \raise1pt\hbox{$>$}}}         

\def\beq{\begin{equation}}
\def\eeq{\end{equation}}
\def\ba{\begin{eqnarray}}
\def\ea{\end{eqnarray}}

\def\<{\langle}
\def\>{\rangle}

\begin{document}

\begin{flushright}
\date{}
\end{flushright}

\title{The Universe in a Soap Film}

\author{Rohit Katti}
\affiliation{Raman Research Institute, Bangalore, India 560 080}
\affiliation{Department of Physics, Bangalore University, Bangalore, 
India 560 056}
\author{Joseph Samuel} 
\affiliation{Raman Research Institute, Bangalore, India 560 080}
\author{Supurna Sinha} 
\affiliation{Raman Research Institute, Bangalore, India 560 080}

\begin{abstract}
The value of the cosmological constant is one of the major 
puzzles of
modern cosmology: it is tiny but nonzero. 
Sorkin predicted, from the Causet 
approach to quantum gravity, that the cosmological constant has 
quantum fluctuations. The predicted order of magnitude of the 
fluctuations agrees with the subsequently 
observed value of the cosmological constant. We had earlier developed an 
analogy between 
the 
cosmological constant of the Universe and the surface tension of fluid 
membranes. Here we demonstrate   
by computer simulations that the surface tension of a fluid
membrane has statistical fluctuations stemming from its discrete
molecular structure. Our analogy enables us to view 
these numerical experiments as probing a small and fluctuating 
cosmological constant. 
Deriving insights from our analogy, we show that a fluctuating 
cosmological constant is a {\it generic} feature of quantum gravity 
models and is far more general than the specific context in which it was originally 
proposed. We pursue and refine the idea of a fluctuating cosmological 
constant and work towards making further testable predictions.
 
\end{abstract}
\pacs{04.60.-m, 82.39.Wj, 87.15.Kg, 87.16.Dg, 46.70.Hg,
68.03.-g}
\maketitle

\section{Introduction}
In the {\it Sand Reckoner}\cite{archimedes}, Archimedes
of Syracuse estimated the number of sand grains that would fill the then 
known Universe. 
Putting the spatial volume of the Universe at somewhat larger than our 
Solar system, he arrived at an astronomical number, which in modern 
notation is
$10^{63}$. This was an early effort at grappling with the large numbers
that are needed to describe our Universe. A modern version of the same
endeavour would replace the {\it spatial} volume of the Universe by its
{\it spacetime} four volume, which is around $10^{112}
\rm{cm^4}$.  The modern analogue of
``a grain of sand'' is the smallest element of spacetime, which from
our theories of relativity ($c$) , gravitation ($G$) and quantum
mechanics ($\hbar$) is around the Planck length $ (\hbar G/c^3)^{1/2}$
scale of $10^{-33}\rm{cm}$.
The known Universe is much larger now than
in Archimedes' time and ``the grains of sand'' much finer. But the
general idea is the same and we can estimate the number of ``grains of
sand'' by dividing the four volume of the Universe by the Planck four
volume. The answer turns out to be around $10^{244}$. We will refer to
this as ``Archimedes' Number'' ${\cal N}_{Arch}$.

{\it Dirac and large numbers:} 
In a somewhat unusual paper, Dirac \cite{Dirac} put forward the
hypothesis that large dimensionless numbers
are unnatural in cosmology and that
one should try to reduce the independent ones by finding relations
between such numbers. In the last two decades, there have been
systematic 
observations\cite{copeland,Perlmutter:1998np,Riess:1998cb,WoodVasey:2007jb}  
of dim and distant supernovae, which
clearly indicate the presence of a tiny (in natural Planck units
$\hbar=c=G=1$) but non zero cosmological
constant $\Lambda\approx 10^{-122}$. Small dimensionless numbers in 
cosmology are as embarrassing as large ones, since the inverse of a 
small number is a 
large one. Can one relate these two large numbers ${\cal N}_{Arch}$
and $\Lambda^{-1}$?
This was precisely what
was done by Sorkin \cite{sorkin90} in a remarkable prediction made {\it
before} the supernova 
observations\cite{copeland,Perlmutter:1998np,Riess:1998cb}.
Sorkin argued that quantum gravity effects would predict an order
of magnitude for fluctuations in the cosmological constant which in
natural units is $1/\sqrt{{\cal N}_{Arch}}$. This is precisely the order
of magnitude of the observed value of the cosmological constant.
Sorkin's proposal was made in the context of Causal Sets, which is
one of several approaches to quantum gravity. We will argue here,
using an analogy developed earlier, that Sorkin's proposal is far more
general than the context in which it was proposed. A {\it fluctuating}
$\Lambda$ is a {\it generic} feature of quantum theories of gravity
and is one of the few predictions which have emerged from them. 
Understanding the fluctuations in more detail may lead to 
testable predictions which could be used to rule out candidate quantum
theories of gravitation.

{\it Approaches to Quantum Gravity:}
While there are many theoretical approaches to quantum gravity,
each with its own adherents, 
few of these approaches have been developed to the point 
where they make significant contact with the real 
world in the form of testable experimental 
predictions.
Some quantum gravity models violate Local Lorentz invariance 
and predict non Lorentz covariant dispersion relations for 
electromagnetic radiation propagating {\it in vacuo}.
There have been proposals\cite{AmelinoCamelia:1997gz} to test such 
models
using gamma ray burst observations. However, not all quantum gravity
models, predict violation of Local Lorentz invariance. 
A more generic feature of quantum gravity models is an underlying 
discreteness of spacetime at the Planck scale.
It is clear from simple dimensional arguments and thought experiments
involving Heisenberg microscopes that spacetime events cannot be 
localised
to better than Planck scale accuracy. 
We learn from semiclassical gravity that black hole entropy is finite 
and given by the number of Planck areas in the horizon.
This is a persuasive argument that spacetime is discrete at the Planck 
scale. Some possible tests of this discreteness have been explored in
\cite{Dowker:2003hb,Philpott:2008vd}.
From this discreteness, we are naturally
led to consider Archimedes' number ${\cal N}_{Arch}$ and look for
possible ways to make predictions which can be tested.

{\it Avogadro's Number and Brownian Motion:}
Let us now turn from large numbers in cosmology to large numbers
in the laboratory. A familiar large number is Avogadro's number
${\cal N}_{Avo}\approx10^{23}$ the number of atoms in a mole of
chemical substance. Atoms are so small and there are so many of them,
even in a grain of sand, that it is hard to discern the
atomistic nature of matter in most experiments. This impasse was broken 
around a hundred
years ago by Einstein's work on Brownian motion. It became
possible
to deduce the existence of atoms by observations of pollen grains
using an optical microscope, a coarse instrument by atomic standards.
Brownian motion is due to {\it fluctuation} effects resulting from
the atomic structure of matter. Such effects are of order
$1/\sqrt{{\cal N}_{Avo}}$ which is not quite as small as
$1/{\cal N}_{Avo}$ and therefore of sensible magnitude.
Could it be that the cosmic acceleration which has been revealed
by supernova observations represents cosmic fluctuations caused by
the otherwise unobservable, quantal (Planck scale) discreteness of
spacetime? Could the accelerating Universe be a
modern analogue of Brownian motion? These are the questions
we address in this paper. We pursue Sorkin's idea and try to understand
it from a general point of view. Our larger objective is to translate
the idea into firm testable predictions and generate a confrontation
between theory and observations.

{\it Analogue Gravity:} Condensed matter analogues of fundamental 
physics
are a valuable aid to understanding\cite{volovik,vachaspati,visser}.
Particle physics has imbibed ideas like spontaneous symmetry breaking
from analogies with magnetic systems and superconductivity.
Historically, the Higgs phenomenon
of elementary particle physics was first noticed
in superconductivity \cite{olesen}.
Such cross fertilization of ideas enriches {\it both} 
fundamental physics \cite{capovilla-2005,capovilla-2002} and its 
laboratory analogue.
Analogue models for gravity are as old as GR itself\cite{visser}.
Laboratory analogues 
\cite{unruh,visser}
of Hawking radiation have been discussed
in the context of supersonic fluid flows and Bose Einstein
condensates. Another
example is the analogy \cite{ajit,vilenkin}
between defects in liquid crystals and phase transitions
in the early universe.
Such analogies provide a concrete context
for discussing fundamental physics and sometimes an
experimental basis for the discussion.
In some domains of fundamental physics, like gravity at the Planck 
scale,
experiments cannot be performed because they are beyond our reach in
energy.
Laboratory analogues are therefore extremely valuable as they are the
nearest one can get to experimental quantum gravity. They
bring abstract ideas of fundamental physics down to earth and
into the laboratory. While laboratory analogues of particle 
physics, classical and semiclassical gravity have been discussed 
earlier, laboratory analogues of quantum gravity effects
are much rarer. Some of these \cite{visser} probe violations of local
Lorentz invariance, which are present in {\it some} quantum gravity 
models. A more generic feature of quantum gravity models is {\it 
discretess} of spacetime in some form, which is present in all quantum
gravity models.
Volovik \cite{volovik} has shown how the Universe can be seen in a Helium droplet. 
We show here how the Universe can be seen in a soap film. In this paper 
we pursue an analogy between quantum gravity and soft condensed matter.
Soft matter systems are dominated by thermal rather than quantum 
fluctuations. Such systems can be studied by table top 
experiments that are relatively easy to perform.

{\it Summary of paper:}
This article is organised as follows. 
We will first review the necessary background, 
describing the cosmological constant
problem, Sorkin's proposed resolution of it and our 
previously developed analogy between surface tension and the 
cosmological constant. We then describe numerical experiments on 
fluid membranes that reveal a small and fluctuating surface tension. 
By analogy this amounts to probing a small and fluctuating cosmological
constant.  
We then use the analogy to arrive at the conclusion that the fluctuating
$\Lambda$ idea is a robust one and would be generically present in any
approach to quantum gravity. We conclude with a discussion of
the limitations of the analogy and some future directions.

\section{Review of Previous Work}
{\it The Cosmological Constant Problem:}
The cosmological constant has long been in disgrace because of its murky 
origins. Einstein introduced it into General Relativity for the wrong 
reasons, then faced with Hubble's observations, recanted and took it out 
(again for the wrong reasons).
For much of the history of GR, the cosmological constant suffered 
benign neglect because of its tainted past.
GR was successfully applied to solar system physics without a 
cosmological constant. In a seminal paper on the cosmological 
constant \cite{Sahni:2008zza}, Zeldovich let the genie out of the bottle 
and it
has never successfully been put back since. 
From quantum field theory we know that
all the fields in nature have vacuum ``zero point'' energy. 
If we sum over all the modes present in  the theory 
we get a divergent answer, which has to be regulated by applying a 
cutoff at the Planck scale. 
This results in a cosmological constant of order one in Planck 
units. In fact, the observed value of the cosmological
constant is tiny but non zero $\Lambda l_{Planck}^4\approx 10^{-122}$.
(Our choice of units for $\Lambda$ is nonstandard but convenient.)
This results in the cosmological constant problem:
\begin{description}
\item a) Why is the cosmological constant so small?
\item b) Why (if it is so small)  does it exist at all?
\end{description}
This is the same problem that faced the learned men of Brobdingnag
when they were confronted with Gulliver and asked to account for his 
small stature: why was he so small? And why (if he was so small) did he
exist at all? It is not easy to solve this problem since the two parts
of it are in opposition. If we succeed in explaining the smallness,
we would be hard put to explain the existence (and vice versa). 
Explaining a small non zero cosmological constant brings us up against Dirac's 
large numbers again.

{\it Sorkin's proposal:} In 1990, Sorkin proposed using the Causet 
approach to quantum gravity that the smallness of $\Lambda$ could be 
understood as a quantum fluctuation.
Sorkin's Causet approach to quantum gravity replaces spacetime by a discrete
structure, a collection of points carrying causal relations. The number
${\cal N}_{Arch}$ of points is the four volume ${\cal V}$ of spacetime
(more precisely the four volume of the causal past of a cosmic observer 
in Planck units).
The rest of the metrical information in spacetime (the conformal structure)
is captured in causal relations between points.

Sorkin's proposal addresses only part b)
of the cosmological constant dilemma. Let us for the
moment suppose that part a) has been solved: some mechanism has been
found for ensuring that the expectation value $<\Lambda>$ of the
cosmological constant is zero. Sorkin's idea
is that there will be Poisson
fluctuations $\Delta \Lambda$
about this mean value which result in
a small nonzero cosmological
constant $\Lambda=<\Lambda>+\Delta \Lambda=\Delta \Lambda$.
From the uncertainty
principle ${\Delta \Lambda} {\Delta} {\cal V}\approx 1$.
From Poisson statistics, $\Delta {\cal N}_{Arch}\approx 
\sqrt{{\cal N}_{Arch}}$. These
$\sqrt{{\cal N}_{Arch}}$ fluctuations are the mechanism for
a small and nonzero
cosmological constant. Based on this argument from quantum gravity,
Sorkin predicts\cite{sorkin90} the following order of magnitude
for the fluctuations in $\Lambda$:
\begin{equation}
\label{lambda}
{\Delta\Lambda}\approx \frac{l^{-4}_{Planck}}{\sqrt{{\cal N}_{Arch}}}
\end{equation}
This gives a logical basis for relating $\Lambda^{-1}$ with the 
Archimedes' number ${\cal N}_{Arch}$ in accordance with Dirac's large
number hypothesis. 

{\it Astronomical Evidence:} Sorkin's prediction is also in accord with 
astronomical data. There is a growing body \cite{copeland} of 
observational evidence 
(e.g redshift-luminosity 
distance relations from type I supernovae) which shows that the
Universe is accelerating at the present epoch,
indicative of a positive cosmological constant. The energy density
in the cosmological constant (referred to as dark energy by astronomers)
is comparable in magnitude to the energy density of matter in the 
Universe. Sorkin's argument predicts the correct
order of magnitude for
$\Lambda$ but makes no prediction about the expected sign of the 
cosmological constant. Neither is there any prediction from Causet 
theory that solves part a). It assumes that part a) is 
solved by some other mechanism and then supplies the small fluctuation 
which would solve part b) and explain the observed value of $\Lambda$ .
Other researchers\cite{paddy,volovik2}
have also followed on Sorkin's idea with slight variations. 

{\it Analogy:}
In \cite{prlsam} we developed an analogy between the Universe 
and a fluid membrane. The analogy is based on the usual mapping between
quantum field theory and statistical mechanics. The geometric 
description of a membrane as a surface in space corresponds to the 
geometric description of spacetime as a four dimensional manifold.
The quantum fluctuations of spacetime map to the thermal fluctuations
of a membrane.
The breakdown of the smooth geometric picture at the molecular scale 
$l_{mol}$ 
corresponds to the Planck scale discreteness of spacetime expected from 
quantum gravity. 
The analogy is explained in Ref.\cite{prlsam} to which the reader is
referred for an explanation of notation and more details.
For the reader's convenience, we have displayed the main points
in tabular form for quick reference (table 1).  
\begin{center}
\begin{table}
\hspace*{0.1cm}{\Large{\bf Table of Analogy}}\\
\begin{tabular}{*{2}{l}}
      \hline
{\bf Membranes} & {\bf Universe}\\
Configuration ${\cal C}$ &  History ${\cal H}$\\
Area of a configuration&  Four volume of a history\\
Sum over configurations&  Sum over histories\\
Energy ${\cal E(C)}$   &  Classical Action ${\cal I(H)}$\\
${\cal E}_0=a_0\int d^2 x \sqrt{\gamma}$ & $I_0=c_0\int d^4 x \sqrt{-g} 
$ 
\\
${\cal E}_2=a_2\int d^2x \sqrt{\gamma}H^2 $ &  $I_2=c_2\int d^4 x 
\sqrt{-g} 
R$\\
Minimum energy configuration  & Classical Path of Least Action\\
Temperature $T$ &  Planck's constant $\hbar$\\
Thermal Fluctuations  & Quantum Fluctuations\\
Surface Tension $\sigma$  & Cosmological Constant $\Lambda$\\*[-0.1cm]
Free Energy  & Effective Action\\*[-0.1cm]
{Molecular Length $l_{\rm mol}=.3{\rm nm}$}&  {Planck Length 
$l_P=10^{-33}{\rm cm}$}\\*[-0.1cm]
{Molecules} &  {Causet elements}\\*[-0.1cm]
{Molecular level discreteness of space ~~~}  & {Planck scale level 
discreteness of space-time}\\*[-0.1cm]
\hline
\end{tabular}
\caption{Table shows the corresponding elements in the analogy 
between the 
Universe and Fluid Membranes}.
\end{table}
\end{center}
It is worth emphasizing 
that the analogy is of a formal and mathematical nature (rather than a 
physical one). As a result we have mappings between physically disparate
elements. For example the analogue of the cosmological constant of the 
Universe is the surface tension of a membrane. 
The {\it statics} of membranes is mapped to the {\it dynamics} of 
spacetime.
What is striking about 
our analogy is  that it maps the exotic and ill understood physics of 
quantum gravity into relatively well understood physics, which is 
testable by laboratory and numerical experiments. 

To illustrate the analogy, we 
show how renormalisation effects generate
a surface tension which is of order $\sigma_0=T/l^2_{mol}$.
Here $T$ is the temperature and $l_{mol}$ a molecular length scale,
which is around nanometres. We will sometimes use ``natural units''
and set $T$ and $l_{mol}$ to unity.
As an instructive example, we 
start with a microscopic energy in which the surface tension is set to 
zero {\it by hand}. Consider such a membrane whose  
equilibrium configuration is a plane 
rectangle with sides $L_1,L_2$ and area ${\cal A}=L_1 L_2$.
Due to thermal fluctuations, the membrane will vibrate about its
equilibrium configuration. Assuming small vibrations, we can model
them as harmonic oscillators and expect by equipartition that 
the expectation value of energy $<E>$ in each mode is $T$. 
Performing a sum over modes to evaluate
the contribution from all the modes we find a divergent answer 
which has to be regulated by the molecular scale cutoff.
\begin{equation}
T\int\int_0^{k_{max}} \frac{d^2k d^2x}{(2\pi)^2}
\label{modesum}
\end{equation}
where ${\vec k}$ is a wave vector and $k_{max}=2\pi/l_{mol}$ is the
cutoff set by the molecular scale. Performing the $k$ integral we find
that this contributes a term
 \begin{equation}
\frac{\pi T}{l^2_{mol}} \int d^2x =\frac{\pi T}{l^2_{mol}} {\cal 
A}
\label{radiative}
\end{equation}
to the energy, giving rise to a surface tension 
$\sigma_0\approx T/l_{mol}^2$ 
of order 1 in dimensionless units. 
Even if one assumes that the microscopic energy
has zero surface tension, such a term is generated by ``radiative 
corrections'' in a manner analogous to the generation  
of vacuum energy from the Casimir effect. So we would expect that 
most membranes have a value of order $\sigma_0$ for the surface
tension. This expectation is borne out for most interfaces. 
However, there is a notable exception -fluid membranes- 
which we now turn to.

{\it fluid membranes:} Fluid membranes\cite{safran} are composed of 
amphiphilic 
molecules, which consist of hydrophilic
(water loving) polar head groups and
hydrophobic (water hating) hydrocarbon tails. Common examples of 
amphiphilic molecules are soaps and detergents, phospholipids, 
cholesterol 
and glycolipids.
If one adds amphiphiles to water, they cluster together to hide their 
tails
and so form supramolecular structures like micelles, vesicles and
symmetric bilayers. Bilayers of phospholipids are studied 
by physicists as simple models for the biological cell membrane. These 
bilayers are sometimes called fluid membranes as the amphiphilic 
molecules flow freely on the two dimensional surface of the membrane.

A fluid membrane, that is not subject to 
external constraints, naturally assumes a tensionless state. 
From dimensional arguments,
we would have expected a surface tension of order 
$\sigma_0$. 
This works out to about $40$ milliJoules$/{\rm m}^2$
in conventional units. 
Indeed, most interfaces have a surface tension which is of this order.
Fluid membranes are an exception in having vanishing surface 
tension. 
A tensionless fluid membrane solves part a) of 
the analogue cosmological constant problem.  
It was shown in \cite{prlsam} that 
the fluid membrane also solves part b) by virtue of its 
small surface tension fluctuations.
This is a direct analog of Sorkin's suggestion \cite{sorkin} that there 
are small fluctuations in the cosmological constant due to spacetime
discreteness on quantum gravity scales within the causet 
framework\cite{PhysRevlett.59.521}. 

{\it The Analogue  Cosmological Constant Problem:} 
From fairly simple theory \cite{piran, safran}
one can understand the
tensionless state of fluid membranes.
As one increases the areal density of molecules in the fluid membrane, 
the area 
per molecule $\alpha$ decreases. There is a limit to this
packing density however, and at a critical value of $\alpha=\alpha_0$,
there is a minimum in the free energy per molecule $f(\alpha)$. 
A fluid membrane which can adjust its areal density does so to attain 
this critical value of $\alpha$ and is said 
to be saturated. At the saturation
point $\alpha=\alpha_0$ the free energy per molecule has a minimum
\begin{equation}
\frac{\partial f}{\partial \alpha}|_{\alpha=\alpha_0}=0
\label{minimum}
\end{equation}
Consider a saturated membrane with area ${\cal A}$   
and $N={\cal A}/\alpha$ molecules. The total free
energy is given by\cite{piran, safran}
$F({\cal A})=N f(\alpha)$. The expected value of the surface
tension of the membrane vanishes:
\begin{equation}
<\sigma>=\frac{\partial F}{\partial {\cal A}}=\frac{\partial f}{\partial
\alpha}|_{\alpha_0}=0.
\label{surfacetension}
\end{equation}
From the analogy developed earlier \cite{prlsam}, this corresponds to 
a vanishing cosmological constant and solves the first horn of the
cosmological constant dilemma.

This simple theory also predicts that a fluid membrane consisting
of $N$ molecules has an
interfacial tension $\sigma$ which fluctuates about zero. The
mean square statistical fluctuation in the surface tension is:
\begin{equation}
(\Delta \sigma)^2=<(\sigma-<\sigma>)^2>= T \frac{\partial^2 F}{\partial
{\cal A}^2}=\frac{T}{N} \frac{\partial^2 f}{\partial\alpha^2}|_{\alpha_0}.
\label{ms}
\end{equation}
On dimensional grounds, we can expect 
$\sigma_0\approx\sqrt{T\frac{\partial^2 
f}{\partial
\alpha^2}|_{\alpha_0}}$ 
and so 
\begin{equation}
\Delta \sigma\sim \frac{\sigma_0}{\sqrt{N}}
\label{rms}
\end{equation}
in complete analogy to Sorkin's proposal in the cosmological context.
From the simple theory described above we would expect that fluid 
membranes which can adjust their areal density (either by exchanging 
lipid molecules with the ambient solution or by adjusting their area) 
a) have vanishing surface tension and b) have 
surface tension fluctuations
about zero with a magnitude $\sigma_0/\sqrt{N}$, where $N$ is the 
number of lipid molecules.

{\it Experiments:} What is the experimental situation? Fluid membranes 
have been 
extensively studied both theoretically and experimentally for many 
years (see Ref. \cite{ipsen} for some references). 
It has long been realised that fluid membranes can 
exist in a 
tensionless state. The effects of the vanishing surface tension can be seen 
in the variety of shapes assumed by red blood cells, which is well 
explained theoretically on the basis of a vanishing surface tension.  
The flickering of red blood cells due to thermal fluctuations has also 
been noted.
However, quantitative experiments do not usually yield a value for this 
small tension but only place upper bounds. For instance, Kwok and 
Evans\cite{kwok} probe the surface tension of a fluid membrane using 
pipette aspiration techniques. The upper bound they give is three 
orders of magnitude smaller than dimensionally expected $ \sigma_0=40 mJ/m^2$. 

The fluctuating surface tension of part b) was first predicted in 
\cite{prlsam} and an experiment was proposed to measure it.
The proposed experiment involved a cylindrical fluid 
membrane in an ambient solution stretched between two tiny rings,
both of radius $r$ in tens of nanometers.
One of the rings is
attached to a piezoelectric translation stage
which can be moved in nanometer steps.
The other ring is attached to a micron sized bead which
is confined in an optical trap.
Fixing the separation $L$ by using a feedback loop,
one can measure the force $F$  on the bead.
This force $F$ is related to the
surface tension by $F=2\pi r \sigma$.
We expect to see fluctuations\cite{prlsam,ensemble} in the surface
tension due to finiteness of $N$
over and above any instrumental noise which may be
present.
Such an experiment can be viewed as an analog
quantum gravity experiment probing a small nonzero fluctuating
cosmological constant. 
There are practical, though not insurmountable, difficulties of working 
with real laboratory fluid 
membranes which are 
small enough to show an appreciable fluctuation. In contrast, a computer 
simulation is cleaner and the relevant tunable parameters are 
more in control. This motivated us to use a computer simulation
or a ``numerical experiment'' to test our theoretical predictions. 

\section{Simulations of Fluid Membranes} 
In Molecular Dynamics Simulations (MDS) \cite{frenkel} one integrates 
the microscopic equations (Newton's equations) 
that govern the motion of each atom. Such 
simulations are computationally expensive 
and therefore not very efficient when one has to study the behaviour
of a system containing a large number of particles for a long time.
A more efficient scheme called Dissipative Particle
Dynamics (DPD) has been developed\cite{hooger,frenkel}, which groups 
atoms into beads and uses a coarse grained description.
The beads interact by 
conservative forces as well as dissipative and random forces. 
These last two
are chosen, consistent with the Fluctuation-Dissipation Theorem,
to ensure that the system equilibriates to a Boltzmann distribution.
It has been shown \cite{groota}
that DPD is orders of magnitude more efficient than MDS and gives the
same results. The increased efficiency of DPD permits us to study 
the surface tension fluctuations of fluid membranes for a range of 
membrane sizes.
The DPD programmes we used were provided to us by M. Venturoli.
More details of these programmes are published in \cite{maddalena}.
The programmes were intended to study biological membranes and have
far more realistic detail and flexibility than we really need for our 
purposes. 
We expect that the broad conclusions of our study do not 
depend on these realistic details.

{\it Simulation Model:}
The system is subject to 
periodic boundary conditions in the $x$, $y$ and $z$ directions so that 
the simulation box has the 
topology of a three-torus. The box has equal extent in the $x$ and $y$
directions and a larger extent (initially) in the $z$ direction. 
The DPD programme 
starts with an initially random distribution  of lipid and water 
molecules. The initial configuration is evolved and the molecules
spontaneously assemble into a bilayer lying in the $x-y$ plane. 
The topology of the 
membrane is a two torus and the geometry of its equilibrium 
configuration is flat intrinsically and extrinsically. The total
number of lipid molecules (counting those on both sides of the 
symmetric lipid  bilayer) 
is varied from $40$ to $400$. A snapshot of a bilayer with $400$ 
molecules is shown in Fig.1.
\begin{figure}
\includegraphics[height=14cm,width=16cm]{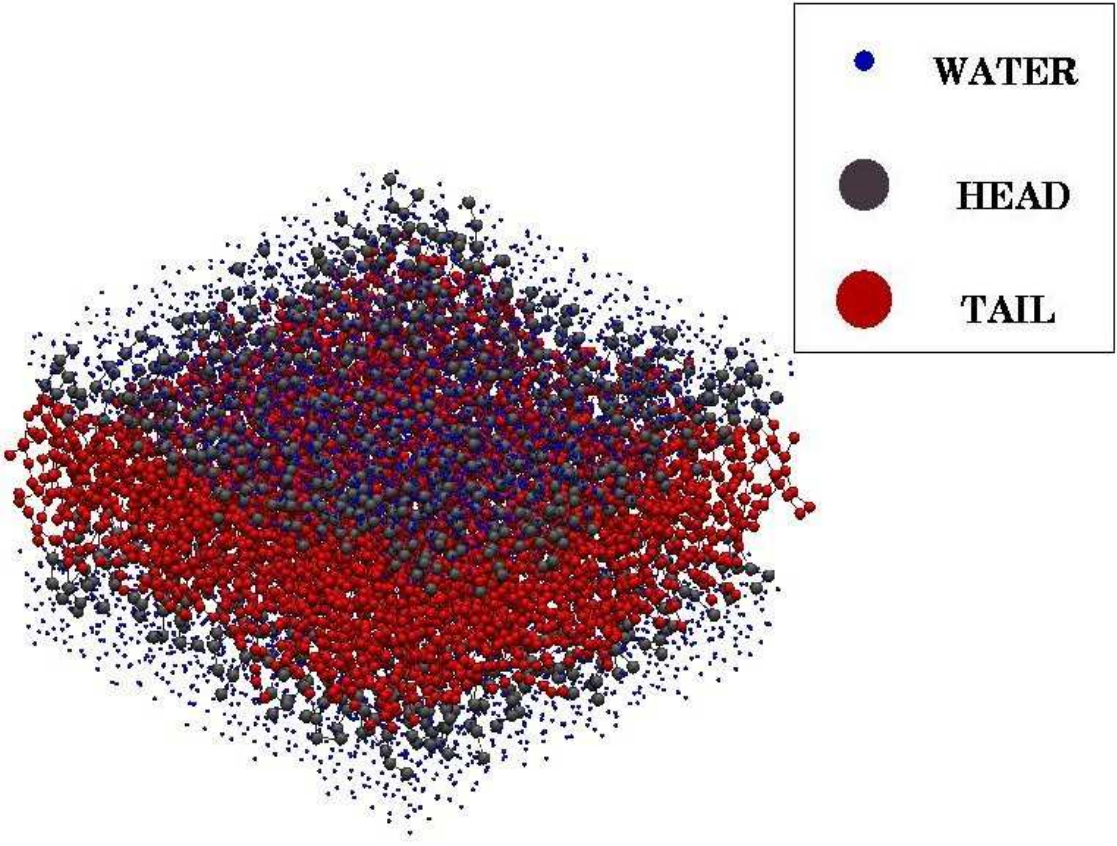}
\caption{The figure shows a system of $400$ lipid molecules forming
a symmetric bilayer with $200$ molecules on each side. The system
is subject to periodic boundary conditions in the $x-y-z$ directions.}
\end{figure}
These numbers have been 
chosen optimally: for small  
numbers of lipid molecules, the bilayer becomes unstable and 
larger numbers are computationally expensive. Such mesoscale membranes 
enable us to probe surface tension fluctuations.

The simulation model\cite{maddalena} uses three types of atoms $w$, 
$h$ and $t$ to represent the 
water, head and tail atoms of a lipid molecule respectively. 
The total force on a DPD particle $i$ is a sum over all the other 
particles $j$, of three pairwise additive type forces:
conservative, dissipative and random.
The 
conservative forces between the particles are all repulsive and given 
by: 
\begin{eqnarray} 
F^C_{ij} &=&a_{ij}(1-|{\vec r}_{ij}|/R_c){\hat r}_{ij} \;\;\;\;{\rm if} 
\;\;
|{\vec r}_{ij}| < R_c\\ 
       &=&0 \;\;\;\;\;{\rm otherwise}
\label{forcea} 
\end{eqnarray} 
$R_c$ is the range of the interaction
and serves as our unit of length. This also determines a unit of time
through the mean thermal velocity\cite{maddalena}. 
The coefficients $a_{ij}>0$ representing the repulsion strength
are suitably chosen to mimic hydrophobic and hydrophilic interactions. 
We use $a=1$ as the unit of energy and the $a_{ij}$ s are chosen as
$a_{ww}=a_{tt}=25$, $a_{wh}=15$, $a_{hh}=35$ and 
$a_{wt}=80$ following \cite{maddalena}. All numbers are expressed in 
in these units.
The temperature $T$ was set to 
$.7$ 
in all the runs. 
The atoms in the lipid molecules are connected via harmonic 
springs with a spring constant $k_r=100$ and equilibrium distance 
$r_0=0.7$. The bond bending potential between two consecutive bonds is 
given in terms of a bending spring constant of $k_{\theta}=10$ and 
equilibrium angle $\theta_0 = 180^{o}$.

{\it Calculating the Surface Tension:}
The surface tension of the membrane has been computed as follows. 
At each point of the membrane, the stress tensor is computed from the 
microscopic forces between pairs of atoms. There is a ``gauge'' 
ambiguity \cite{schofield} in associating a stress field with a force 
between two 
atoms.
This ambiguity arises because our description does not include a
mechanism for interaction between atoms. (The van der Waal interaction  
is fundamentally electromagnetic and would entail distributed Maxwell 
stresses.) A reasonable and widespread 
choice \cite{schofield,maddalena} is to suppose that the 
stress is concentrated along
the shortest straight line joining each pair of atoms. 
After summing over all pairs, we arrive at the total stress tensor 
$\sigma_{ij}$.

From symmetry, the stress tensor is 
expected to have principal directions normal and tangential to the 
membrane. The normal component is $p_N(z)$ and the two lateral 
components are $p_L(z)$. Both these pressures have an intricate
variation across the membrane. The surface tension was computed from the
formula\cite{maddalena}:
\begin{equation}
\sigma=\int dz (p_N(z)-p_L(z))
\label{surfacetensioncalc}
\end{equation}
where the integral is over the thickness of the membrane. This 
integral was evaluated numerically and the result
was averaged over $100$ runs. 
Each run consisted of evolving the 
system for $10,000$ timesteps (where each timestep translated into 
real units is of the order of $25$ picoseconds).    
This amounts to 
time averaging the surface tension. We may believe from the ergodic
theorem that this is equivalent to performing an ensemble average.

The formula (\ref{surfacetensioncalc}) for the surface tension 
is free of the ``gauge'' ambiguities that appear in the stress field.
It  can be readily understood by 
considering a spherical 
membrane and computing the pressure difference between the inside and
outside of the membrane by integrating the equilibrium 
condition $\nabla_i \sigma^{ij}=0$, for the stress tensor 
$\sigma^{ij}$
across the membrane, and using Laplace's law relating the surface 
tension to the pressure 
difference and the extrinsic curvature of the membrane.

A real bilayer membrane which is not subject to external constraints
is expected to assume a tensionless state. However, bilayers assembled via
simulations are constrained by the finite size of the simulation box and 
this induces lateral tensions on the membrane, so it takes some effort
to attain a tensionless state\cite{goetz,lipo}. With $N$ fixed, we vary 
the lateral size of the simulation box and thus vary $\alpha$.
In Fig.2
we plot the membrane tension versus $\alpha$, the area per lipid for 
different numbers of molecules. From the simple theory explained above,
we would expect that the surface tension $\sigma=f(\alpha)$ depends only 
on $\alpha$ and
doesn't depend on $N$. This is precisely what was seen in the 
simulations. To a very good approximation, the graphs for different $N$ 
values fall on top of each  other.
This enables us to locate the area per molecule pertaining 
to the tensionless state. 
(We are being a bit sloppy here. The $\alpha$ plotted above
is the {\it projected} area per lipid, which is equal to the
area per lipid only in the equilibrium state.)
We repeat this for a number of different values of 
total number of molecules. We notice that at a preferred value of 
the area per molecule 
$\alpha_0$ (around $1.57$ in all cases) the membrane is tensionless.
(See Fig. $2$).  This is entirely in agreement
with the standard picture for tensionless membranes. If we allow  
the simulation box to adjust itself and ``relax'' under the pull 
of the lipids, the membrane naturally assumes a tensionless state.
This is the analogue of the vanishing cosmological constant of the Universe
and solves part a).

\begin{figure}
\includegraphics[height=8.2cm,width=10cm]{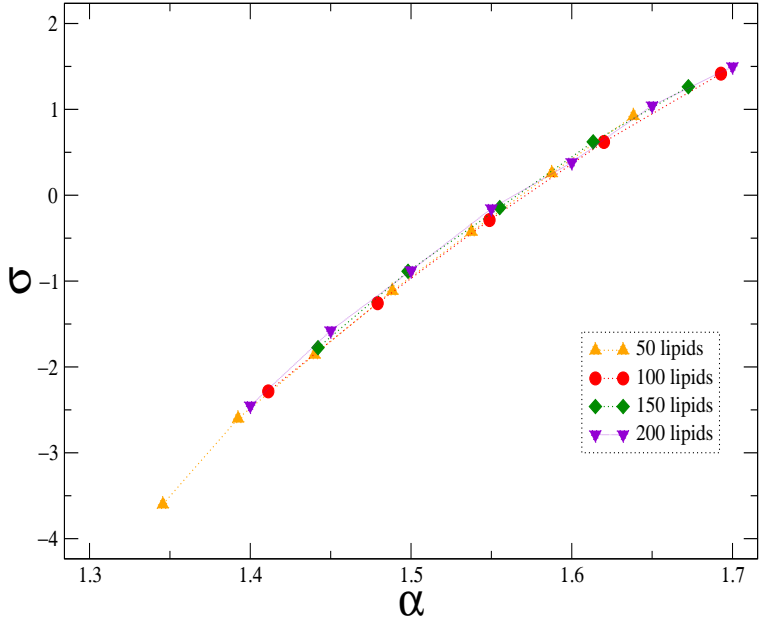}
\caption{The surface tension $\sigma$ plotted against the 
area per molecule $\alpha$ for various values of $N$, the number of 
molecules constituting the membrane.}
\end{figure}
Regarding part b) (the fluctuating surface tension), in Fig. 3, we plot 
the standard deviation $\Delta \sigma$ 
of surface tension as a function of $N$, the 
number of lipids. The log-log 
plot clearly 
shows the $1/\sqrt{N}$ dependence of the surface tension on $N$. As N 
tends to infinity, we expect to find vanishing surface tension 
fluctuations in the thermodynamic limit.  
The coefficient of $1/\sqrt{N}$ agrees in order of magnitude with 
$\sigma_0$ as expected. This numerical experiment can be viewed as a 
demonstration of Sorkin's idea of a small and fluctuating cosmological
constant.
\begin{figure}
\includegraphics[height=7.0cm,width=10cm]{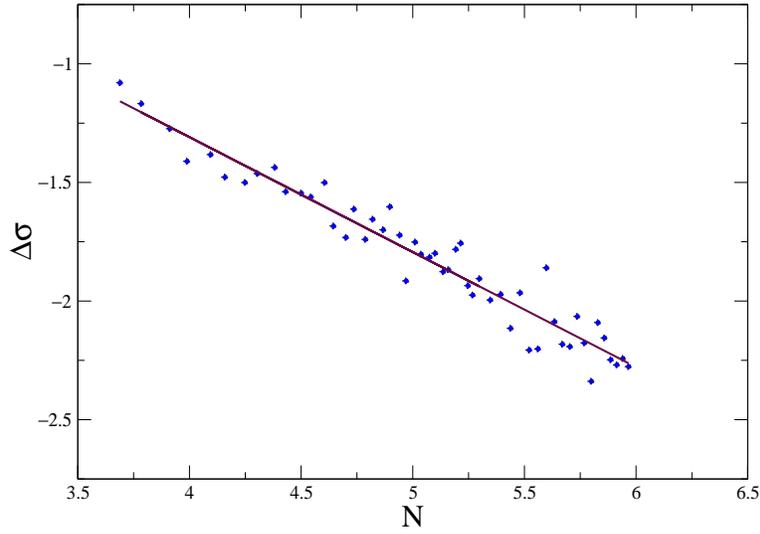}
\caption{The mean square fluctuations in the surface tension 
$\Delta\sigma$ plotted against 
the number $N$ of molecules constituting the membrane on a log-log plot.
The best fit straight line to the data has a slope of $.48$,
very close to the theoretically expected slope of $.5$.}
\end{figure}

The fluctuation effects we see in the membrane simulations are entirely 
standard predictions from Statistical Mechanics \cite{landau}. Consider 
a system 
described by a set of variables $X$, whose entropy is given 
by $S(X)$. A thermodynamic system maximises its entropy subject 
to the constraints imposed on it. This determines the equilibrium value
of $X$, which we call $X_0$. Since $X_0$ is a maximum of the entropy
the first derivatives of $S(X)$ vanish at $X_0$. Expanding $S(X)$ in 
a Taylor expansion about $X_0$, we see that changes in entropy can be 
approximated by a quadratic form: the Hessian matrix of $S$ with respect
to $X$. As first discussed by Einstein, a system in statistical 
mechanics explores neighboring lower entropy states with 
probability given by 
$\exp{\Delta S}$.
Thus there are Gaussian fluctuations of $X$ 
about $X_0$ with a width determined by the eigenvalues of the Hessian
matrix. The fluctuations responsible for Brownian motion are exactly
of this variety. The equilibrium position of our membrane is 
a flat surface in the $x-y$ plane with an area determined by the 
periodic boundary conditions. However, due to thermal fluctuations
the membrane explores curved configurations, leading to area 
as well as surface tension fluctuations. While the fluctuation effects 
seen and their $N$ dependence is to be expected, they have not so far 
been experimentally 
studied.  
Neither has their theoretical significance for quantum gravity
been appreciated. This is what we turn 
to next.

\section{Learning from Analogy}
Sorkin's suggestion of a fluctuating $\Lambda$ neatly solves the 
Brobdingnag problem of explaining a small but non zero cosmological 
constant: $\Lambda$ is as small as it can be, given the uncertainties 
inherent in quantum gravity. As a bonus, it also automatically solves
the ``coincidence problem''. The measured value for $\Lambda$ today is 
comparable to $\rho$, the matter density in the Universe. According to 
the standard picture of a cosmological ${\it constant}$, 
$\Lambda$ must have been 
subdominant in the past
and is only {\it now} starting to dominate the dynamics of the universe.
But ``why {\it now}?'', a puzzle known as the coincidence problem. It 
appears  to be a coincidence that we 
live in the epoch where $\Lambda$ and $\rho$ are of comparable 
magnitude. In Sorkin's proposal, there is nothing special about our 
present epoch. $\Lambda\approx \rho$ at all epochs, in four
spacetime dimensions. This conclusion of an ``everpresent 
$\Lambda$'' is unaffected by extra dimensions \cite{Sorkin:2005fu} 
provided only that these are 
not {\it large} extra dimensions, much bigger than the Planck 
scale. Sorkin's 
argument implies that 
the order of 
magnitude of the fluctuations in $\Lambda$ was larger in the past than 
it 
is today.
A common reaction from astronomers is that such a theory can be 
observationally ruled out. This reaction is usually based on the 
standard notion of cosmological {\it constant} which is not a 
fluctuating quantity. If indeed, Sorkin's idea can be disproved 
observationally, it would be wonderful, for then, we would have a real 
confrontation between theory and experiment in quantum gravity. 
The logic behind Sorkin's suggestion is simple enough to be 
convincing. If a simple chain of logic leads to a conflict with 
observation (remember Olber's paradox) there is much to be learned since
one of the links in the chain must give. 
However, the idea of a fluctuating cosmological constant is still 
in its infancy  and needs development and precision  
to make further testable predictions.  A number of 
points remain unclear:
What does one mean by a fluctuating cosmological constant?
Is it a fixed fluctuation for the whole universe? Does $\Lambda$ 
depend on cosmic epoch? or cosmic location? Does this not conflict 
with Einstein's equations which imply a constant $\Lambda$ via the 
Bianchi identity?
What is the role of Einstein's equations? How general is Sorkin's 
idea? Is it inherently tied to the Causet approach to quantum gravity,
or do other approaches also predict a fluctuating $\Lambda$?

Let us now learn from the soft matter analogy and address these 
questions. The first lesson we learn concerns our attitude
towards spacetime, gravity and the cosmological constant. 
The orthodox view of spacetime is that of a smooth manifold. This is a 
picture that works well over a range of scales. 
Gravity is normally
viewed by relativists as a  fundamental theory of {\it pure} 
geometry,  and if a cosmological constant is present, $\Lambda$ is 
viewed as a fixed number whose numerical value is to be determined from 
experiment. However, borrowing a leaf from condensed matter physics,
we see that the 
geometric description of a membrane as a smooth surface embedded in 
space is only an emergent picture valid at long (say micron) length 
scales. At a molecular scale (nanoscale) a membrane is roughened 
\cite{goetz} by relative displacements and protrusions of individual 
molecules jostling for space. The surface tension emerges at low 
energies as a ``chemical 
potential'' describing the energy cost of creating a unit area of 
membrane. Like any chemical potential, the surface tension can fluctuate
for finite systems as we have seen it does. In Gravity, 
we should by analogy, view GR as an emergent effective description
valid at long length scales. 
There may be a true microscopic theory
(similar to the molecular description of membranes) underlying GR, but 
we do not know what it is and 
are not here attempting to guess what it is. 
Whatever the microscopic 
description is (String Theory, Loop Quantum Gravity, Causets, 
Non-commutative Geometry or any other candidate), we restrict ourselves
to long wavelength phenomena and try to find low energy 
imprints of Planck 
scale effects, just as Brownian motion was manifested at micron scales.
In this spirit the cosmological constant should be viewed as a
``chemical potential'' for the creation of spacetime. It is the action 
cost for creating a unit four volume of spacetime, just as the surface 
tension is the energy cost for creating a unit area of membrane. 
Like any chemical potential, the cosmological constant too is subject to 
fluctuations.

What does one mean by a fluctuating cosmological constant?
From the membrane analogy, we know that for a flat membrane in its
equilibrium configuration, the surface tension is a constant. This 
follows from the condition for mechanical equilibrium
\begin{equation}
\nabla_i\sigma^{ij}=0
\label{equilibrium}
\end{equation}
However this equation only holds in the thermally averaged sense. As we 
saw earlier, a finite lipid membrane
has fluctuations about its equilibrium configuration. We would thus
expect that the surface tension has local fluctuations which depend on 
position and have vanishing mean: 
\begin{equation}
<\sigma(x)>=0\\,
<\sigma(x)\sigma(x')>\neq0
\label{spectrumsigma}
\end{equation}
We would similarly expect in the cosmological context that
$\Lambda$ has spacetime dependent fluctuations.
\begin{equation}
<\Lambda(x)>=0\\,
<\Lambda(x)\Lambda(x')>\neq0
\label{spectrumlambda}
\end{equation}
We see already from Sorkin's argument (since it can be applied at 
any epoch) that the fluctuating $\Lambda$ is not a fixed constant for 
the whole Universe: it does depend on Cosmic time. However, then it 
would be natural to permit it to have spatial fluctuations as well as temporal
ones. We conclude that the $\Lambda$ fluctuations do depend on Cosmic
epoch and location.

{\it Einstein's Equations:} Einstein's equations imply 
via the Bianchi identity that $\Lambda$ must be a constant. 
Yet we have seen that 
$\Lambda$ has local fluctuations. How does one reconcile the two? 
The membrane analogy suggests an answer. The condition for mechanical
equilibrium (\ref{equilibrium}) is only true on thermal averaging. 
It is not true as an operator equation. Similarly, in cosmology,
we are considering quantum fluctuations about a 
classical solution (the homogeneous, isotropic Robertson-Walker 
metric). Quantum 
fluctuations do not obey the classical Einstein equations. We would 
still expect 
Einstein's equations to hold in some average $<G_{\mu \nu}>=<T_{\mu 
\nu}>$ sense, but there is no reason to expect then to hold as operator 
equations. Similarly the conservation of matter implied by 
Einstein's equations classically, $\nabla_\mu 
T^{\mu \nu} 
= 0$ only holds on an average and not as an operator identity.
It follows that a fluctuating $\Lambda$ does not have to be a constant
in space and time as naively implied by Einstein's equations.

Sorkin's proposal for a fluctuating $\Lambda$ was made within the 
context of Causal Sets and 
Unimodular gravity. How essential are these inputs? Let us consider
them in turn.\\
{\it Causal Sets} 
In Causal Set theory, spacetime is approximated
by a {\it sprinkling} of discrete points which are Poisson distributed
as a result of Local Lorentz invariance. The $\sqrt{N}$ fluctuations
have their origin in the Poisson fluctuations of $N$. A Poisson 
distribution has the property that its variance $<(N-<N>)^2> $is 
equal to its mean $<N>$ for all $N$. 
Note however, that the lipid membrane system has a 
distribution of lipids which is
certainly not Poisson. There is repulsion between the 
molecules and their positions are anticorrelated. However, the system 
still shows  $\sqrt{N}$ fluctuations to a good approximation. 
The $\sqrt{N}$ fluctuations seen in the 
bilayer are a result of the central limit theorem and the law
of large numbers: for large enough $N$, any distribution tends to a 
Gaussian distribution whose width increases as $\sqrt{N}$.
It appears then that Poisson distribution is not essential 
for ${\sqrt{N}}$ fluctuations, if $N$ is large enough.
In the cosmological context, ${\cal N}_{Arch}$ is comfortably large and
we can safely suppose that there will be $\sqrt{{\cal N}_{Arch}}$ 
fluctuations of
the cosmological constant. It appears then that the essential 
ingredient of Sorkin's argument is {\it discreteness} of spacetime,
rather than any other aspect of Causet theory (like Local Lorentz 
invariance).

{\it Unimodular Gravity:} What is the role played by Unimodular gravity?
To understand this we go back to the membrane system.
The surface tension is a thermodynamic 
potential conjugate to the area of the membrane just as pressure
is conjugate to volume. Let us define the Gibbs ensemble
partition function $Z_G(\sigma)$ 
\begin{equation}
Z_{G}(\sigma) = \sum_{{\cal C}} (\exp{- \beta {\cal E}_{2}({\cal C})}) 
(\exp{-\sigma {\cal A}({\cal C})})
\label{gibbs}
\end{equation}
in which the surface tension is regarded as the control 
parameter and the area 
fluctuates. It is also possible to 
define the ``Helmholtz'' ensemble partition function 
\begin{equation}
Z_{F}({\cal A}) = \sum_{{\cal C}} \exp - \beta {\cal 
E}_{2}({\cal C})
\delta({\cal A} - {\cal A}({\cal C}))
\label{helm}
\end{equation}
which describes a system in which the area is held constant and
the surface tension fluctuates. The Helmholtz partition function
relates to the Gibbs partition function by a Laplace transform:
\begin{equation}
Z_{G}(\sigma) = \exp - \beta G(\sigma) = \int_0^\infty d{\cal A}\;\;
e^{-\beta{\sigma{\cal A}}}\;\;e^{-\beta{F({\cal A})}}.
\end{equation}
For systems at the thermodynamic limit the fluctuations are negligible 
and the two ensembles are  equivalent, but for the finite 
size systems that we are interested in, there are 
differences\cite{ensemble}. Which ensemble is appropriate depends
on the experimental conditions. If the control parameter is the 
surface tension then the Gibbs ensemble should be used. One can also
envisage more general ensembles in which neither the area nor the 
surface tension is held constant and both fluctuate. The fluctuations
are related by something like an ``uncertainty'' relation $\Delta {\cal 
A} \Delta {\sigma}\approx 1 $. 
Indeed this is the situation in the simulations we described in the last 
section. All we are holding constant in the simulation is the cyclic boundary 
conditions on the molecules. The area of the membrane {\it and} the
surface tension show fluctuations of order $\sqrt{N}$ and $1/\sqrt{N}$
respectively in accord with the ``uncertainty'' relation.

Using the analogy above we can transcribe the description of a fluid
membrane to the gravity context. We can write analogue ``partition 
functions''. In both cases (membranes as well as 
gravity) we are 
assuming that the microscopic physics provides an ultraviolet cutoff so 
that the sums in (\ref{gibbs},\ref{helm},\ref{gibbsgrav},\ref{helmgrav}) 
are convergent. 
The ``Gibbs Partition function'', which we write formally as
\begin{equation}
Z_G[\Lambda]:=\sum_{\cal H} \exp{i(\Lambda *{\cal V}+I_2)}
\label{gibbsgrav}
\end{equation}
can be viewed as simply one statistical ensemble in which $\Lambda$ is 
held fixed and its conjugate variable ${\cal V}$ fluctuates. 
One could 
equally well consider the conjugate Helmholtz ensemble
\begin{equation}
Z_F[{\cal V}]:=\sum_{\cal H} \exp{(i I_2)} \delta({\cal V}({\cal 
H})-{\cal V})
\label{helmgrav}
\end{equation}
$I_2$ here is the Einstein-Hilbert action (Table 1).
If one takes the classical limit and evaluates this path integral 
by stationary phase one ends up with a theory in which the variations
of the metric in the Einstein-Hilbert action are subject to the 
constraint that the four volume is unchanged. This theory is equivalent
to Unimodular Gravity! Thus Unimodular gravity 
\cite{unruh,unruhwald} should 
not be viewed as
an exotic theory, but a close cousin of Einstein's general relativity
belonging to a different statistical ensemble and related to it 
by a Legendre transform. Classically the two theories are equivalent. 
In fact one can conceive of more general ensembles where {\it both} 
$\Lambda$ and $\cal V$ are allowed to fluctuate. This is no different
from having states in quantum mechanics which are neither position nor
momentum eigenstates. There are fluctuations in both quantities
and these fluctuations satisfy the uncertainty principle.

It would then seem that the two essential components to Sorkin's idea
of a fluctuating cosmological constant are graininess of spacetime 
and a possibility of fluctuations in $\Lambda$. The first 
ingredient is present in most 
approaches to quantum gravity. The second ingredient is easily 
incorporated by considering a conjugate ensemble. We conclude then
that the idea of a fluctuating cosmological constant is not
closely tied up with Causets and Unimodular gravity, but is more
general. This conclusion is disappointing for Causets, because the 
cosmological constant then does not favour Causets over competing 
theories of quantum gravity. On the other hand, the conclusion is
a generic prediction of quantum gravity models and so Sorkin's idea 
gives us a natural resolution of the cosmological constant problem 
from quantum gravity.
\section{conclusion}
As we remarked earlier, the analogy between the Universe 
and fluid membranes is a mathematical one and maps physically 
disparate elements into each other. This sometimes has unexpected 
benefits. The cosmological constant problem is widely regarded as a 
difficult one. However, it appears that the analogue problem is 
almost trivially solved. Free standing fluid membranes have a vanishing 
surface tension (part a) and have small surface tension 
fluctuations (part b). While the vanishing surface tension of fluid 
membranes is regarded as commonplace in 
the soft matter community, its analogue in cosmology is far from 
trivial. It is worth spending a few words on why freestanding fluid 
membranes have vanishing surface tension.
The hydrophobic interactions which govern the behaviour of lipid 
membranes are of high energy ($a_{ij}>>T$). Such strong interactions
give rise to strong forces that regroup particles (in the form of a 
bilayer) so that these strong forces are no longer effective. 
All we see at low energies is the residual effects that remain 
after these strong forces cancel out. These residual effects are a 
consequence of imperfect cancellations between the strong forces.
Similar physics happens in other systems.
Strong electromagnetic forces between protons and electrons 
are screened out when they form atoms and then interact by weak van der 
Waal forces. This idea was expressed early and very well by Newton 
(Opticks): ``Now the smallest particles of Matter may cohere by the 
strongest Attractions, and compose bigger Particles of weaker Virtue; 
and many of these may cohere and compose bigger Particles whose Virtue 
is still weaker, and so on for divers Successions, until the Progression 
end in the biggest particles on which the Operations in Chymistry, and 
the Colours of natural Bodies depend, and which by cohering compose 
bodies of a sensible magnitude.''
Replacing the archaic ``Virtue'' by its modern equivalents 
``charge'' or ``coupling constants'' we see that 
Newton has in mind ``emergent behaviour''. The idea that physics depends 
on scale is a prevalent one in condensed matter physics and formalised
in the renormalisation group.

Our attitude of viewing the supposedly fundamental theory of 
general relativity as an emergent description valid at low energies 
is borrowed from condensed matter physics and is fairly 
widespread today \cite{volovik,hu}. In Landau theory, one 
expands the free  energy in powers of decreasing length dimension and 
discards higher order terms. 
Landau theory is essentially a low energy theory and has the
advantage that we do not need to know the microscopic basis for the 
theory. 
The low energy description depends only on general symmetry 
properties and not on details of the microscopic theory. Our ignorance
of microscopics is absorbed into a set of phenomenological constants.
For instance altering the values of $a_{ij}$ which describe the forces
between atoms  will not change the $N$ dependence of 
$\sigma_0/\sqrt{N}$, but could affect the value of $\sigma_0$.

For quantum gravity, the microscopic theory is presently a 
matter of speculation and therefore it is of advantage not to have to 
commit 
ourselves to any particular model. Doing effective field theory with 
gravitation is no harder than doing continuum elasticity, 
even though matter is fundamentally atomistic. Our ignorance of 
fundamental physics can be lumped into a few ``elastic constants''
of which the cosmological constant is one.
We must distinguish here between {\it generic} predictions
which do not depend on the microscopic theory and {\it specific}
predictions which do. Both have 
their value: generic predictions give us general quantum gravity
insights into cosmology but do not discriminate between different
microscopic theories, while specific predictions can hopefully 
be used to rule out some theories and favour others. For instance, we 
have seen that
the $1/\sqrt{{\cal N}_{Arch}}$ dependence of the $\Lambda$ 
fluctuations is generic. But the coefficient which is of order unity
may depend on the specific microscopic theory.

We have relied on analogy as a guide to refine Sorkin's idea of a
fluctuating cosmological constant. 
Like all analogies, our analogy between GR and membranes
has its limitations (see table 2).
\begin{center}
\begin{table}
\hspace*{1.5cm}{\large{\bf Limitations of Analogy}}\\
\begin{tabular}{*{2}{l}}
      \hline
{\bf Membranes} & {\bf Universe}\\
dimension two  &  dimension four\\
Euclidean geometries&  Lorentzian geometries\\
Positive $\sigma$ minimises area &  Positive $\lambda$ 
causes accelerated expansion\\
No Causal Structure& Causal Structure\\
Boundary value problem & Initial value problem\\
Ambient Space and Extrinsic geometry&  Purely Intrinsic 
geometry\\
Exponentially damped sum over configurations& Oscillatory 
phase sum over histories\\
Non-Poissonian distribution of molecules&Poissonian distribution
of Causet elements \\
\hline
\end{tabular}
\caption{Table 2 shows some limitations of the analogy between General 
Relativity  and fluid membranes}
\end{table}
\end{center}
Obvious differences are those of dimension
(four versus two) and signature (Lorentzian versus Euclidean).
As a result of these differences,
a positive $\Lambda$ tends to accelerate the expansion of the Universe,
whereas a positive $\sigma$ tends to contract a membrane!
The Action of spacetime depends on purely intrinsic geometry,
while the membrane energy depends on both intrinsic and
extrinsic geometry.  There is no 
analogue in membranes of the Causal structure of spacetime which was
a prime motivating force in Sorkin's approach. The membrane problem is
Euclidean and
best formulated as a boundary value problem, while in GR because of
the Lorentzian signature, an initial value problem is more natural.
Despite these differences, the 
fluctuations expected in Sorkin's approach appear in membranes and 
lead us to believe that Sorkin's idea is truly more general than 
its original context. 
With all its limitations, the analogy is a useful one and 
suggests future work in several directions.

Sorkin's idea of a  fluctuating lambda is still incompletely explored.
We feel it is a compelling idea which should be pursued to its logical 
conclusion and confronted with observational evidence. Barrow 
\cite{barrow} has argued
that observational Cosmic Microwave Background Radiation (CMBR) 
evidence rules out the idea of ``Everpresent $\Lambda$'' in the form it 
was developed by Ahmed et al \cite{ahmed}. But this avatar 
of Sorkin's idea\cite{Sorkin:2007bd}
holds fast to some components of Einstein's equations, while giving up 
others. Barrow's criticism too has recourse to Einstein's equations. 
It would be nice to develop Sorkin's idea in its full generality
using stochastic differential equations to replace Einstein's 
equations  and {\it then} try to rule it out convincingly by using CMBR
obervations and improved supernova data which may become available
in the future \cite{Howell:2009mt}

One observable quantity which could be used to confront theory with
experiment is the {\it spectrum} of fluctuation. 
The power spectrum of surface tension fluctuations is easily explored 
using computer techniques as well as analytical calculations\cite{leibler}.
This information is contained in the Fourier transform of the two point
correlation functions $<\sigma(x)\sigma(x')>$.
Since $\Lambda$ has spacetime dependent fluctuations, one could expect 
that  its spectrum of fluctuations \cite{Volovik:2008jf} can also be 
calculated from $<\Lambda(x)\Lambda(x')>$. Since 
the spectrum of metric fluctuations is well constrained by CMBR 
observations, this could provide a confrontation between theory and
experiment. There may be  
aspects of the spectrum that are specific and depend on the 
microscopic theory, in which case the observations
may be used to rule out candidate quantum theories of 
gravitation. On the other hand if the predictions are generic 
and independent of the microscopic theory, a conflict with 
observations would cast a shadow of doubt on the whole venture of 
quantum gravity.

The $\Lambda$ fluctuations may be accessible by numerical quantum 
gravity,
a developing area in which there is considerable current interest.
There is much work \cite{ambjorn} that explores quantum gravity on a 
computer using dynamical triangulations and spin networks. 
Just as we are able to see fluctuating
$\sigma$ in fluid membranes by numerical experiments, it should be
possible to see fluctuating $\Lambda$ in numerical quantum gravity. 
Apart from the magnitude of the effect, it should also be possible to
work out the {\it spectrum} of fluctuations, which is an imporant
observable quantity. Our work suggests that 
by using computer methods like dynamical triangulations it should be 
possible to see  effects of fluctuating $\Lambda$ in every approach to 
quantum gravity. It is also  perhaps true that the simulation methods 
are more highly developed in condensed matter physics
than in numerical quantum gravity.
Techniques analogous to DPD which improve the efficiency 
of computer runs may be a useful import into numerical quantum gravity.

The smooth picture we have of spacetime in GR may yield in quantum 
gravity to some discrete microscopic structure. 
Such views about geometry and spacetime were expressed very early by
Riemann and Einstein (see \cite{gomb}). Riemann in his inaugural
lecture on geometry discusses continuous manifolds as well as 
discrete ones, evidently regarding the latter as more natural.
``..Their quantitative comparison 
happens for discrete manifolds through
counting, for continuous one through measurement.''
Einstein, in a letter to to D\"allenbach, (1916) says
``...But you have correctly grasped the drawback that the continuum
brings. If the
molecular view of matter is the appropriate one, i.e., if a
part of the Universe
is to be represented by a finite number of moving points, then the
continuum
of the present theory contains too great a manifold of possibilities. I
also believe
that this too great is responsible for the fact that our present means
of description miscarry with the quantum theory. The problem seems to me
how one can formulate statements about a discontinuum without calling
upon a continuum...''

It may be as Riemann and Einstein anticipated that 
the spacetime continuum emerges only in 
the low energy approximation 
that is accessible to us
through our feeble probes. Our best particle accelerators today 
are but coarse instruments by Planck scale standards, reminiscent of 
optical microscopes in nanophysics.  
The underlying discreteness of 
spacetime resists our best efforts, just as atoms escaped detection  
for two millenia from the time of Democritus.
If they are numerous enough, discrete points 
(atoms of matter or of spacetime) can appear continuous as the stars on 
the Milky Way. To get at the underlying discreteness, our best chance may be 
to look at emergent {\it fluctuation} effects. This is the main lesson 
we have learned from Einstein's work on Brownian motion and the 
main point of this paper. We hope that the bridge we have made between quantum gravitational physics
and membrane physics will trigger further work in experiments, simulations and theory in both areas.


{\it Acknowledgements:}
It is a pleasure to thank Maddalena Venturoli for 
her simulation programs for the surface tension of lipid membranes,
Abhishek Choudhury for many discussions on this work, 
Irina Pushkina for discussions on numerical quantum gravity, David 
Finkelstein for a conversation about the analogy and Joe Henson, Satya 
Majumdar,
John Ipsen, Michael Fisher, Bei-Lok Hu and Ted Jacobson for their comments.

%

\end{document}